\documentclass[twocolumn,letter]{jpsj3} 
\title{Possible Superconducting Symmetry and Magnetic Correlations in K$_{0.8}$Fe$_2$Se$_2$: \\ A $^{77}$Se--NMR Study}

\author{Hisashi \textsc{Kotegawa}$^{1,3}$\thanks{E-mail address: kotegawa@crystal.kobe-u.ac.jp}, Yudai \textsc{Hara}$^{1}$, Hiroki \textsc{Nohara}$^{1}$, Hideki \textsc{Tou}$^{1,3}$, \\ Yoshikazu \textsc{Mizuguchi}$^{2,3}$, Hiroyuki \textsc{Takeya}$^{2,3}$, and Yoshihiko \textsc{Takano}$^{2,3}$}

\inst{$^{1}$Department of Physics, Kobe University, Kobe 658-8530, Japan\\
$^{2}$National Institute for Materials Science, 1-2-1, Sengen, Tsukuba 305-0047, Japan\\
$^{3}$JST, Transformative Research-Project on Iron Pnictides (TRIP), Chiyoda, Tokyo 102-0075, Japan\\

}

\abst{
We report $^{77}$Se-NMR measurements in the recent discovered K$_{0.8}$Fe$_2$Se$_2$ with the superconducting transition temperature $T_c=32$ K.
The characteristic point of this system is that the high density of state is located in the vicinity of the Fermi level.
The comparison between Knight shift and nuclear spin lattice relaxation rate $1/T_1$ shows that the antiferromagnetic spin correlation is not strong in this system, but it is reasonable to consider that its development toward low temperatures occurs.
$1/T_1$ in the superconducting state can be reproduced well by an $s^{\pm}$-wave model, but a $d$-wave model also roughly reproduces the data.
Non exponential behavior in $1/T_1$ at low temperatures disagrees with a single isotropic gap.

}

\kword{K$_{0.8}$Fe$_2$Se$_2$, superconductivity, NMR}

\begin{document}
\maketitle

The recent discovery of superconductivity in K$_{0.8}$Fe$_2$Se$_2$ is accelerating the research on related materials.\cite{Guo,Krzton}
The research will spread widely because of the relatively high transition temperature $T_c$ of $\sim32$ K and absence of arsenic.
Another interesting point is whether superconductivity occurs in the same framework as that of other Fe-based superconductors.
Most Fe-based superconductors possess both hole-like bands and electron-like bands.
The nesting between those bands is considered to yield multigap superconductivity characterized by $s^{\pm}$ symmetry.\cite{Mazin}
The band calculations suggest that  the electronic state of stoichiometric KFe$_2$Se$_2$ is far from that of other Fe-based superconductors, and the appearance of the common band topology to other systems depends on the amount of hole doping, which is probably induced by the deficiency of K and Fe.\cite{Shein,Cao,Nekrasov}
Recent angle-resolved photoemission spectroscopy (ARPES) studies suggest that K$_{0.8}$Fe$_2$Se$_2$ is a heavily electron-doped system compared with other Fe-based superconductors, and the hole-like Fermi surface disappears.\cite{Qian,Zhang}
If superconductivity is realized under such a situation, $s^{\pm}$ symmetry should be excluded.

In this letter, we report $^{77}$Se-NMR results in single-crystalline K$_{0.8}$Fe$_2$Se$_2$ to investigate the superconducting (SC) symmetry and magnetic correlations.

A single-crystalline sample with $T_c=32$ K was prepared as described elsewhere.\cite{Mizuguchi}
$^{77}$Se-NMR measurement using a standard spin-echo method was performed under a magnetic field of $\sim8.995$ T along the $ab$ plane in both the normal state and the SC state.
$^{77}$Se possesses the nuclear spin of $I=1/2$, which corresponds to one NMR transition.
$T_c=31$ K was estimated under $\sim8.995$ T from the onset of diamagnetism.
The Knight shift was obtained using the gyromagnetic ratio of $\gamma_n=8.13$ MHz/T.
The nuclear spin-lattice relaxation rate $1/T_1$ was obtained by a nice fitting of the recovery curve to the single exponential function in the normal state.
In the SC state, we omitted a small amount of fast relaxation arising from the vortex core in the fitting.

\begin{figure}[htb]
\centering
\includegraphics[width=0.9\linewidth]{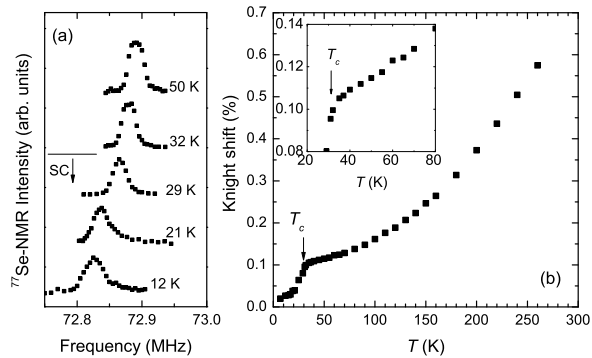}
\caption[]{(a) $^{77}$Se-NMR spectrum under $\sim8.955$ T at several temperatures. The spectrum shifts both below and above $T_c$. (b) Temperature dependence of Knight shift. The strong temperature dependence at high temperatures originates in the band structure effect. Knight shift drastically decreases below $T_c$.}
\end{figure}

Figure 1(a) shows the $^{77}$Se-NMR spectra measured by applying $\sim8.955$ T along the $ab$ plane.
The line width was 25 kHz at $32$ K.
The spectral shape at high temperatures is not the Lorenz type but rather close to a rectangle, suggesting that slight inhomogeneity exists in the sample.
However, we confirmed that the electronic state is homogeneous, within the accuracy of $1/T_1$, from the frequency dependence of $1/T_1$.
Knight shift was estimated from a central position of the spectrum in the normal state.
The spectrum shows a slight and asymmetric broadening owing to the existence of a vortex in the SC state, where Knight shift was obtained from the peak of the spectrum.
The spectrum at 29 K just below $T_c$ shifts from that at 32 K without obvious broadening, ensuring that $T_c$ is almost homogenous in the sample.
Figure 1(b) shows the temperature dependence of Knight shift.
In the normal state, Knight shift shows a strong temperature dependence.
This reduction in Knight shift toward low temperatures originates in the high density of states (DOS) near the Fermi level,\cite{Ikeda} which is often seen in the electron-doped systems among Fe-based superconductors.\cite{Nakai,Grafe,Ning,Imai}
In K$_{0.8}$Fe$_2$Se$_2$, Knight shift continues to decrease down to $T_c$ without the Fermi liquid region, as shown in the inset.
This indicates that the high DOS is located in the vicinity of the Fermi level, and the spin susceptibility of $q=0$ is suppressed with decreasing temperature.
Knight shift markedly decreases below $T_c=31$ K and approaches $\sim0.02$\% at zero temperature.
Generally, Knight shift consists of the temperature-dependent spin part $K_s$ and temperature-independent orbital part (or chemical shift) $K_{orb}$.
This suggests that the spin part of superconducting symmetry is a singlet and $K_{orb} \sim 0.02$\%, because field-induced DOS is not expected at the present magnetic field sufficiently lower than $H_{c2}$.\cite{Mizuguchi}

\begin{figure}[htb]
\centering
\includegraphics[width=0.8\linewidth]{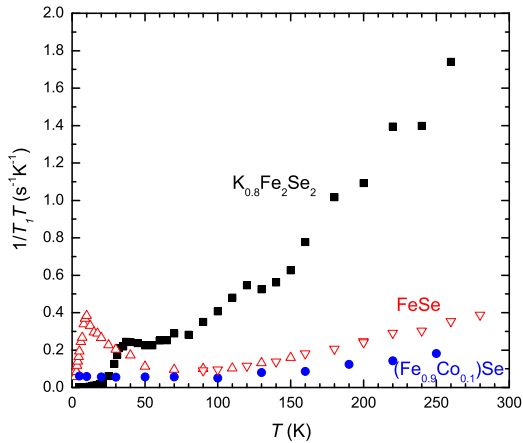}
\caption[]{(color online) Temperature dependences of $1/T_1T$ for K$_{0.8}$Fe$_2$Se$_2$, FeSe, and non superconducting (Fe$_{0.9}$Co$_{0.1}$)Se. $1/T_1T$ also shows a strong temperature dependence at high temperatures, but it is almost constant below $\sim60$ K. The large difference from the related systems suggests that the high DOS near the Fermi level is characteristic feature of K$_{0.8}$Fe$_2$Se$_2$. In FeSe, the difference in the symbols indicates the difference in the sample. Low-temperature data were obtained using a high-quality sample, and there is no sample dependence at high temperatures.}
\end{figure}

Figure 2 shows temperature dependences of $1/T_1T$ for K$_{0.8}$Fe$_2$Se$_2$, FeSe, and non superconducting (Fe$_{0.9}$Co$_{0.1}$)Se.
In K$_{0.8}$Fe$_2$Se$_2$, $1/T_1T$ also decreases with decreasing temperature owing to the band structure effect.
The marked difference from Knight shift is that $1/T_1T$ is almost constant below $\sim60$ K, while Knight shift continues to decrease toward low temperatures.
This disagreement suggests that the spin fluctuation of $q\neq0$, which is most likely an antiferromagnetic (AF) one, develops at least in this temperature region, even though it is not so strong.
A strong temperature dependence of DOS toward low temperatures likely masks the development of spin fluctuations.
We compare $1/T_1T$ with those in FeSe and electron-doped (Fe$_{0.9}$Co$_{0.1}$)Se.
$1/T_1T$ in FeSe was measured using a high-quality sample made with high-temperature annealing,\cite{Mizuguchi_review} and the NMR line width of which is almost the same as that of stoichiometric FeSe.\cite{Imai}
Even in this sample, $1/T_1T$ separated into two components below 40 K, so that we plotted the short component of $\sim75$\% in the volume fraction, which is considered to be intrinsic.
In FeSe, an obvious increase in $1/T_1T$ is seen below $\sim40$ K owing to the development of a low-energy part of AF spin fluctuations.
The development of $1/T_1T$ is enhanced under pressure with a close relationship with $T_c$.\cite{Imai,Masaki}
In the high-temperature region, the temperature dependence of $1/T_1T$ in K$_{0.8}$Fe$_2$Se$_2$ is markedly stronger than that in FeSe, and it is weak in (Fe$_{0.9}$Co$_{0.1}$)Se.
The reduction of $1/T_1T$ in (Fe$_{0.9}$Co$_{0.1}$)Se upon electron doping indicates that the high DOS, which is located below the Fermi level in FeSe, is removed from the Fermi level by electron doping.
This doping suppresses the AF spin fluctuations and superconductivity.
By contrast, strong temperature dependence in $1/T_1T$ above $\sim60$ K in K$_{0.8}$Fe$_2$Se$_2$ suggests that high DOS is located closer to the Fermi level than in the case of FeSe.
However, the development of AF spin fluctuations at low temperatures is not drastically enhanced in K$_{0.8}$Fe$_2$Se$_2$.
This indicates that the character of the band near the Fermi level is different between FeSe and K$_{0.8}$Fe$_2$Se$_2$.

\begin{figure}[htb]
\centering
\includegraphics[width=0.8\linewidth]{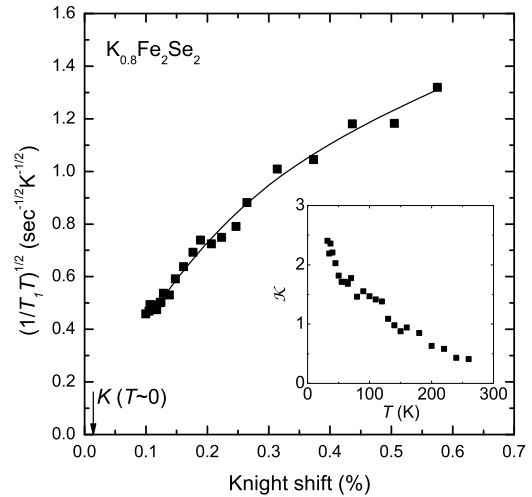}
\caption[]{Relationship between Knight shift and $(1/T_1T)^{1/2}$ from 32 to 260 K. The non linear relationship indicates the collapse of the Korringa relation. The solid curve is a guide for the eye. The arrow indicates Knight shift at the lowest temperature, which is considered to correspond to $K_{orb}$. The inset shows the temperature dependence of Korringa ratio \texttt{K} estimated using $K_{orb}=0.02$\%.}
\end{figure}

Figure 3 shows Knight shift vs $(1/T_1T)^{1/2}$ in the normal state.
The Korringa ratio \texttt{K} is given as follows.

\begin{equation}
\texttt{K}=\frac{1}{T_1TK_s^2}\frac{\hbar}{4\pi k_B} \frac{\gamma_e^2}{\gamma_n^2}
\end{equation}
Here, $\gamma_e$ is the electron gyromagnetic ratio.
\texttt{K} corresponds to the character of spin correlations, for example, $\texttt{K}<<1$ corresponds to ferromagnetic correlations and $\texttt{K}>>1$ corresponds to AF correlations.
The observed strong temperature dependences of Knight shift and $1/T_1T$ are mainly attributed to the band structure effect, but Knight shift vs $(1/T_1T)^{1/2}$ does not show a linear relationship, indicating that the Korringa relation of $\texttt{K}=const.$ is not realized.
The relationship of upward convex suggests that \texttt{K} increases with decreasing temperature, because the slope in the figure is proportional to $\texttt{K}^{\ 1/2}$.
If we assume $K_{orb}=0.02$\% as $K_s(T=0)=0$, the temperature dependence of \texttt{K} is as indicated in the inset.
\texttt{K} is not very far from 1, indicating that spin correlations are not strong in K$_{0.8}$Fe$_2$Se$_2$, but the obvious increase in \texttt{K} toward low temperatures suggests the development of AF spin correlations.

\begin{figure}[htb]
\centering
\includegraphics[width=0.7\linewidth]{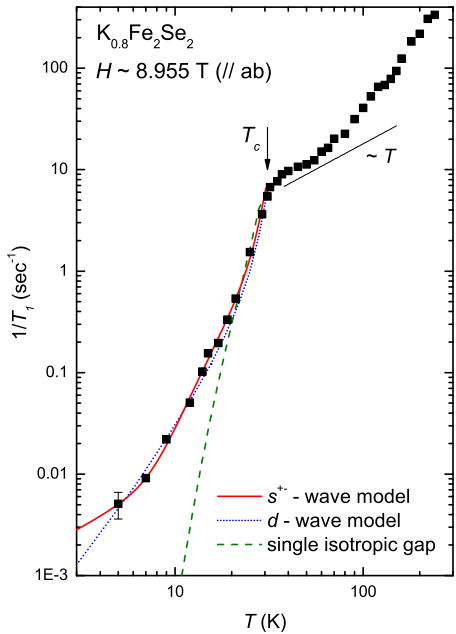}
\caption[]{(color online) Temperature dependence of $1/T_1$ for K$_{0.8}$Fe$_2$Se$_2$. $1/T_1$ does not show a coherence peak just below $T_c$. Non exponential behavior excludes a single isotropic gap. The red solid curve are the best-fitting result obtained using the $s^{\pm}$-wave model. The blue dotted curve (the green dashed curve) was obtained using the $d$-wave model (a single isotropic model).}
\end{figure}

Next we focus on the symmetry of the superconducting gap in K$_{0.8}$Fe$_2$Se$_2$.
Figure 4 shows the temperature dependence of $1/T_1$.
The steep decrease in $1/T_1$ was observed below $T_c$ without any sign of a coherence peak.
The power is $T^6$-like just below $T_c$, approaches $T^3$-like below $\sim T_c/2$, and then starts to deviate from $T^3$ below $\sim T_c/5$.
Such power-law behavior in $1/T_1$ is often seen in most Fe-based superconductors.
The absence of the coherence peak and non exponential temperature dependence of $1/T_1$ clearly excludes the possibility of a single isotropic gap pointed out from the ARPES measurement.\cite{Zhang}
The green dashed curve is a calculation using a single isotropic gap of $\Delta_0=4k_BT_c$ ignoring the coherence effect.
This curve completely disagrees with the data for the low-temperature region.
Non exponential behavior suggests that the multigap is realized with different magnitudes of gap size or that the anisotropic superconductivity is realized with a finite DOS in the SC gap.
The red solid curve is the results of a calculation based on the $s^{\pm}$ isotropic gap.\cite{Nagai}
This model has been adapted for other Fe-based systems such as LaFeAsO, (Ba,K)Fe$_2$As$_2$, and LiFeAs.\cite{Matano,Yashima,Li}
We can reproduce the data well using $\Delta_1=4.2 k_BT_c$, $\Delta_2=1.7 k_BT_c$, $N_1:N_2=0.7:0.3$, and $\eta=0.048\Delta_1$.
Here, $\Delta_1$ and $\Delta_2$ are the magnitudes of the SC gap in the respective bands, and $N_1$ and $N_2$ are respective DOS at the Fermi level.
$\eta$ is the smearing factor due to the impurity effect, and induces the residual DOS at the Fermi level.
$\Delta_1=4.2 k_BT_c$ indicates the superconductivity in the strong-coupling regime, and is comparable to that of (Ba$_{0.6}$K$_{0.4}$)Fe$_2$As$_2$.\cite{Yashima}
The value of $\eta$ is also comparable to those of other Fe-based superconductors with $s^{\pm}$ symmetry.\cite{Matano,Yashima,Li}
The multigap feature is not obvious in the Knight shift probably due to the relatively large $N_1$, which is similar to the case of (Ba$_{0.6}$K$_{0.4}$)Fe$_2$As$_2$.\cite{Yashima}
The third model is a two-dimensional $d$-wave model ($\Delta(\theta)=\Delta_0 \cos(2\theta)$) with line nodes that produce $T^3$ behavior at low temperatures.
As shown by the blue dotted curve, this model also can reproduce the data roughly using $\Delta_0 = 5 k_BT_c$ in the strong-coupling regime and the residual DOS of 3\%, which is generally induced by impurity scattering.
The model of $d$-wave symmetry is not sufficiently far from the data to exclude the possibility that $d$-wave symmetry is realized in this system.
We tried the two-gap $d$-wave model, but it did not improve the fitted result.
The $1/T_1$ at the lowest temperature gives the upper limit of the residual DOS of $\sim7$\%.
The small residual DOS of $\sim3-7$\% indicates that impurity scattering is small in spite of the system with a deficiency of ions.
The $s^{\pm}$ symmetry is more plausible, but for its identification, measurement at lower temperatures will be required, as well as under lower field to avoid the contribution in the relaxation process from the vortex core.

The ARPES measurement suggests the isotropic gap opens only in the electron pocket.\cite{Zhang}
The interpretation based on our data seems to be inconsistent with this observation.
Another ARPES measurement also shows that this system possesses only an electron pocket.\cite{Qian}
If the nesting between the electron pockets plays a crucial role in superconductivity, the $d$-wave symmetry is likely to be realized.
The present NMR result cannot exclude this possibility.
From the results of our study, we can claim that the single isotropic gap is excluded, and the $s^{\pm}$-wave or $d$-wave with a strong-coupling regime are possible candidates.

Finally, we mention earlier NMR reports presented by Yu and co-workers quite recently.\cite{Yu,Ma}
The obtained data in this study are similar to their data.
The difference in the absolute value of Knight shift seems to originate in the different value of $\gamma_n$.
Yu and co-workers claim that AF spin correlations are absent in this system from the Korringa ratio.
Note that we defined the reciprocal of their Korringa ratio as \texttt{K}.
The \texttt{K} is sensitive to the estimation of $K_{orb}$, and they estimated $K_{orb}$ considering that Korringa ratio is constant against temperature. (they use the notation of $K_{c}$).
However, our measurement at higher temperatures shows that the Korringa ratio depends on temperature.
If we can obtain the susceptibility data removing the contribution from impurity, it will help us to estimate the exact $K_{orb}$.
As for the symmetry of the SC gap, they tried the fitting of $1/T_1$ with an isotropic gap and discussed the possibility of the multigap.
We clearly excluded a single isotropic gap and suggest the multigap or the anisotropic gap from the fitting of $1/T_1$.

In summary, we performed NMR measurements in K$_{0.8}$Fe$_2$Se$_2$ using a single-crystalline sample.
The strong temperature dependences in Knight shift and $1/T_1T$ at high temperatures suggest that the high DOS is located in the vicinity of the Fermi level.
This feature is much stronger than that in FeSe.
The breakdown of the Korringa relation suggests the AF spin correlations develop toward low temperatures, although it is not obvious compared with FeSe.
The temperature dependence of $1/T_1$ in the SC state can exclude the single isotropic gap.
It can be reproduced well by the $s^{\pm}$-wave model, but the possibility of $d$-wave symmetry still exists.

This work has been partly supported by Grants-in-Aid for Scientific Research (Nos. 19105006, 20740197, 20102005, 22013011, and 22710231) from the Ministry of Education, Culture, Sports, Science, and Technology (MEXT) of Japan.

\end{document}